\documentclass[doublecol]{epl2}

\title{Sudden death and robustness of quantum discord and entanglement in cavity QED}
\shorttitle{Sudden death and robustness of quantum discord and entanglement} 

\author{J. S. Zhang \inst{1,*} \and L. Chen \inst{1} \and M. Abdel-Aty \inst{3} \and A. X. Chen$^{1,4}$}
\shortauthor{J. S. Zhang \etal}

\institute{
  \inst{1} Department of Applied Physics, East China Jiaotong University,
Nanchang 330013, People's Republic of China
\\
  \inst{2} Centre for Quantum Technologies, National University of Singapore,  Singapore 117542 \\
  \inst{3} Mathematics Department, Faculty of Science, Sohag University, 82524
Sohag, Egypt \\
  \inst{4} Centre for Atom Optics and Ultrafast Spectroscopy, Swinburne University of Technology, Melbourne 3122, Australia\\
  $^*$Corresponding author: jszhang1981@zju.edu.cn
}

\abstract{The quantum dynamics of two entangled two-level atoms is studied.
Each of the two atoms is located within an isolated and dissipative cavity.
If the interaction time of atoms and cavities is not very long,
the amount of quantum discord and entanglement between two atoms decreases
as the system evolves.
The sudden death of quantum discord and entanglement of two atoms occurs within
a short interaction time.
However, after a long interaction time,
quantum discord and entanglement of two atoms could be partially preserved
due to the long-lived nature of quantum discord and entanglement.
Surprisingly, we find the amount of long-lived quantum discord could be smaller than
that of long-lived entanglement.
Thus, entanglement may be more robust than quantum discord against decoherence.}

\begin{document}

\maketitle

\section{Introduction}

Quantum entanglement is at the heart of quantum information processing and
quantum computation \cite{Nielsen2000,Raimond2001,Chen2010}. In recent years,
many efforts have been invested in the study of the evolution of joint
systems formed by two subsystems (each system locally interacts with its environment)
 \cite{Yu2004,Yu2009,Bellomo2008,Rau2008,Zhang2009,Jamroz2006}. In particular,
the entanglement of a two-qubit system
may disappear for a finite time during the dynamics evolution. The nonsmooth finite-time disappearance
of entanglement is called \textquotedblleft entanglement sudden
death\textquotedblright\ (ESD). Experimentally, the ESD phenomenon has been
observed in the laboratory by several groups for optical setups \cite{Almeida2007,Salles2008}
and atomic ensembles \cite{Laurat2007}.

One the other hand, quantum entanglement is not the only kind of
quantum correlation useful for quantum information processing \cite{Bennett1999,Horodecki2005,Niset2006}.
In fact, it was shown both theoretically \cite{Braunstein1999,Meyer2000,Datta2005,Datta2007,Datta2008,Dillenschneider2008,Sarandy2009,Cui2010}
and experimentally \cite{Lanyon2008} that some tasks can be sped up over their classical counterparts using
fully separable and highly mixed states. These results clearly show that
separable states with quantum discord can be used
to implement quantum information processing such as deterministic quantum computation with
one qubit \cite{Lanyon2008}.
Quantum discord introduced in \cite{Ollivier2001,Henderson2001}
is another kind of quantum correlation different from entanglement.
Very recently, quantum discord has been investigated widely \cite{Wang2010,Auyuanet2010,Sun2010,Werlang2009}.
Note that all the previous studies \cite{Wang2010,Auyuanet2010,Sun2010,Werlang2009}
have shown that, for several quantum
systems, there is no quantum discord sudden death (DSD).
However, quantum discord is a kind of quantum correlation in composite quantum systems.
Since entanglement of quantum systems can stay zero for a finite time (ESD), a natural
question is whether there is DSD in quantum systems with ESD.
Here, we present a quantum system where there exists DSD as well
as ESD. We also explain why there is no DSD in \cite{Werlang2009}.
Furthermore, we find that there is also long-lived quantum discord.
In recent years, many efforts has been devoted to the study of
the long-lived entanglement in cavity QED \cite{Yu2006,Xu2005,Aty2006,Dajka2007} or
solid state systems \cite{Aty2008}.
To the best of our knowledge, there is few study on
the long-term behavior of quantum discord. Thus, an investigation of
quantum discord of a quantum system in the presence of decoherence
in the limit $t\rightarrow\infty$ is highly desired.
This question is also addressed in the present work.

In the present paper, we investigate the dynamics of quantum discord and entanglement
 of a quantum system formed by two two-level atoms within
two spatially separated and dissipative cavities in the dispersive
limit using the results of \cite{Luo2008}.
The two atoms are initially prepared in the Werner states \cite{Bellomo2008} and the cavities are initially
prepared in coherent states.
We show that both DSD and ESD can appear
in the present system. The amount of quantum discord and entanglement of two atoms decreases
with time in the short-term. However, the long-term behavior is
very different since a long survival of quantum discord and entanglement are shown in
the system. This implies that
 quantum discord and entanglement of two atoms
could be partially preserved even they are put into dissipative
cavities.
Unlike the results in \cite{Wang2010,Werlang2009},
our results show that the amount of long-lived quantum discord could be smaller
than that of long-lived entanglement. In other words, quantum entanglement may
be more robust than quantum discord in the present model.

\section{The model}
We first consider a quantum system consisting of a two-level atom
interacting with a single-mode cavity. Under the electric dipole and
rotating wave approximation, the Hamiltonian of the present system is $%
(\hbar =1)$ \cite{Scully1997}
\begin{eqnarray}
H=\omega a^{\dag }a+\frac{\omega _{0}}{2}\sigma _{z}+g(a^{\dag }\sigma
_{-}+a\sigma _{+}),
\end{eqnarray}
where $g$ is the atom-field coupling constant, $\sigma _{\pm }$ are the atomic spin
flip operators characterizing the effective two-level atom with frequency $%
\omega _{0}$, and $\sigma _{z}=|e\rangle
\langle e|-|g\rangle \langle g|$. Note that the symbols
$|e\rangle $ and $|g\rangle $ refer to the excited
and ground states for the two-level atom.
Here, $a^{\dag }$ and $a$ are the
creation and annihilation operators of the field with frequency $\omega $,
respectively. The dispersive limit is obtained when the condition $|\Delta
|=|\omega _{0}-\omega |\gg \sqrt{n+1}g$ is satisfied for any relevant $n$.
Then, the interaction Hamiltonian $g(a^{\dag }\sigma _{-}+a\sigma _{+})$ can
be regarded as a small perturbation. Hence the effective Hamiltonian of the
present model can be rewritten as \cite{Meystre1992}
\begin{eqnarray}
H_{e}=\omega a^{\dag }a+\frac{\omega _{0}}{2}\sigma _{z}+\Omega \lbrack
(a^{\dag }a+1)|e\rangle \langle e|-a^{\dag }a|g\rangle \langle g|],\nonumber\\
\end{eqnarray}
with $\Omega =g^{2}/\Delta $. In the interaction picture, the interaction
Hamiltonian is
\begin{eqnarray}
V=\Omega \lbrack (a^{\dag }a+1)|e\rangle \langle e|-a^{\dag }a|g\rangle
\langle g|].
\end{eqnarray}

We assume the two-level atom interacting with a coherent field in a
dissipative environment. This interaction causes the losses in the cavity
which is presented by the superoperator $\mathcal{D}=\gamma(2a\cdot a^{\dag
}-a^{\dag }a\cdot -\cdot a^{\dag }a)$, where $\gamma$ is the decay constant. For
the sake of simplicity, we confine our consideration in the case of zero
temperature cavity. Then, the master equation that governs the dynamics of
the system can be written as follows
\begin{eqnarray}
\frac{d\widetilde{\rho}}{dt}=-i[V,\widetilde{\rho}]+\mathcal{D}\widetilde{\rho},  \label{master}
\end{eqnarray}%
where $\widetilde{\rho}$ is the density matrix of the atom-field system.
If the initial state of the two-level atom is $\left(
\begin{array}{cc}
\zeta_{a} & \zeta_{c} \\
\zeta_{c}^{\ast } & \zeta_{b}%
\end{array}%
\right) $ and the field is initially prepared in a coherent state $|\alpha
\rangle =e^{-|\alpha |^{2}/2}\sum_{n=0}^{\infty }\frac{\alpha ^{n}}{\sqrt{n!}%
}|n\rangle $ with $\alpha $ being a complex number. Here, $|n\rangle $ is
the Fock state with $a^{\dag }a|n\rangle =n|n\rangle $.
Then, the reduced density matrix of the atom is obtained by tracing out the
variables of the field from the atom-field density matrix \cite{Zhang2010}
\begin{eqnarray}
\widetilde{\rho} _{atom}(t) &=&\zeta_{a}|e\rangle \langle e|+\zeta_{b}|g\rangle
\langle g| +[\zeta_{c}f(t)|e\rangle \langle g|+h.c],  \nonumber \\
f(t)&=&\exp {\{-i\Omega t+|\alpha |^{2}(e^{-2\gamma t}-1)\}}  \nonumber \\
&& \times \exp {\{\frac{|\alpha |^{2}\gamma}{\gamma+i\Omega }[1-e^{-2(\gamma+i\Omega )t}]\}} \nonumber\\
&& \times \exp\{|\alpha|^2e^{-2\gamma t}(e^{-2i\Omega t}-1)\},  \label{s1}
\end{eqnarray}%
where $h.c$ denotes the Hermitian conjugate.

Then, we consider a quantum system consisting of two
noninteracting atoms each locally interacts with its own coherent field of
a dissipative cavity.
The interactions between each atom and its own dissipative cavity is described by Eq. (\ref{master}).
We assume the two atoms are initially prepared in Werner states
defined by \cite{Bellomo2008}
\begin{eqnarray}
\rho _{\Phi } &=&p|\Phi \rangle \langle \Phi |+\frac{1-p}{4}I,  \nonumber \\
\rho _{\Psi } &=&p|\Psi \rangle \langle \Psi |+\frac{1-p}{4}I,  \nonumber \\
|\Phi \rangle  &=&\frac{1}{\sqrt{2}} (|eg\rangle +|ge\rangle),  \nonumber \\
|\Psi \rangle  &=&\frac{1}{\sqrt{2}} (|ee\rangle +|gg\rangle) ,\label{wl}
\end{eqnarray}
where $p$ is a real number which indicates the purity of initial states, $I$
is a $4\times 4$ identity matrix. The parameter $p$ is 1 for pure sates and 0 for
completely mixed states.
The two fields are prepared in coherent states $|\alpha _{1}\rangle $
and $|\alpha _{2}\rangle $. For the sake of simplicity, we assume $\alpha
_{1}=\alpha _{2}=\alpha$, the decay rates of the two cavities are equal, and the
atom-field coupling constants are the same. Using the method introduced in \cite{Bellomo2008},
we can obtain the reduced density matrix of two atoms
conveniently.

The reduced density matrix of two atoms can be obtained by
using the superoperator method \cite{Zhang2010,Zhang20091,Zhang20092}.
As one can see below, the quantum discord and entanglement of the present system
can be calculated conveniently 
by employing the results of \cite{Luo2008}.
We assume the initial state of two atoms is $\rho_{\phi}$.
Using the results of \cite{Zhang2010} and Eq. (\ref{wl}),
we obtain the density matrix of two atoms $\rho(t)$ as follow
\begin{eqnarray}
\left(
\begin{array}{cccc}
\frac{1-p}{4} & 0 & 0 & 0 \\
0 & \frac{1+p}{4} & \frac{p|f(t)|^2}{2} & 0 \\
0 & \frac{p|f(t)|^2}{2} & \frac{1+p}{4} & 0 \\
0 & 0 & 0 & \frac{1-p}{4}
\end{array}%
\right),\label{densitymatrix}
\end{eqnarray}
where $f(t)$ is given by Eq. (\ref{s1}).

\section{Quantum discord and entanglement}
In general, a composite quantum system contains both quantum and classical correlations,
the total amount of which are quantified by quantum mutual information. Precisely, the quantum
mutual information of a composite bipartite system $\rho^{AB}$ is defined as
\begin{eqnarray}
\mathcal{I}(\rho^{AB})&=&S(\rho^A)+S(\rho^B)-S(\rho^{AB}),\label{qmi1}
\end{eqnarray}
where $S(\rho)=-Tr(\rho \log_2{\rho})=-\sum_i(\lambda_i\log_2{\lambda_i})$
is the von Neumann entropy of density matrix $\rho$ with $\lambda_i$ being the eigenvalues of density
matrix $\rho$. We note that $0\log_2 0$ is defined to be 0 and
$\rho^A$($\rho^B$) is the reduced density matrix of $\rho^{AB}$ by tracing out system $B(A)$.

Quantum discord \cite{Ollivier2001,Henderson2001} is another kind of quantum correlation different from entanglement.
In order to quantify quantum discord, the authors of \cite{Ollivier2001} proposed
to use the von Neumann type measurements consisting of one-dimensional projector $\{\mathcal{B}_i\}$
(acts on system $B$ only),
such that $\sum_i\mathcal{B}_i=1$. The conditional density matrix of the total system
 after the von Neumann type measurements is \cite{Ollivier2001}
\begin{eqnarray}
\rho^{AB}_{\mathcal{B}_i}&=&\frac{1}{p_i}(I\otimes\mathcal{B}_i )\rho^{AB}(I\otimes\mathcal{B}_i ),\nonumber\\
p_i&=&Tr((I\otimes\mathcal{B}_i )\rho^{AB}(I\otimes\mathcal{B}_i )),
\end{eqnarray}
where $p_i$ is the probability of the corresponding measurement.
The quantum conditional entropy with respect to this kind of measurement
is defined as
\begin{eqnarray}
S(\rho^{AB}{|\{\mathcal{B}_i\}})=\sum_i p_i S(\rho^{AB}_{\mathcal{B}_i}), \label{qce1}
\end{eqnarray}
and the corresponding
quantum mutual information with respect to the measurement is defined by
\begin{eqnarray}
\mathcal{I}(\rho^{AB}|\{\mathcal{B}_i\})=S(\rho^A)-S(\rho^{AB}|\{\mathcal{B}_i\}).
\end{eqnarray}
The quantity $\mathcal{I}(\rho^{AB}|\{\mathcal{B}_i\})$ is the information gained about system $A$
if one performs measurement $\mathcal{B}_i$ on system $B$.
The resulting classical correlation according to \cite{Ollivier2001,Henderson2001}
is defined as
\begin{eqnarray}
\mathcal{J}(\rho^{AB})&=&\sup_{\{\mathcal{B}_i\}}\mathcal{I}(\rho^{AB}|\{\mathcal{B}_i\})\nonumber\\
&=&S(\rho^A)-\min_{\{\mathcal{B}_i\}} [S(\rho^{AB}|\{\mathcal{B}_i\})]. \label{cc1}
\end{eqnarray}
The quantum discord is obtained by subtracting $\mathcal{J}$ from the quantum mutual information $\mathcal{I}$
\begin{eqnarray}
\mathcal{D}(\rho^{AB})=\mathcal{I}(\rho^{AB})-\mathcal{J}(\rho^{AB}). \label{qd1}
\end{eqnarray}
As one can see from the above equations,
the minimization procedure should be done over all possible von Neumann measurements $\mathcal{B}_i$
on system $B$. Thus, the main difficulty of calculating quantum discord lies
in the elaborate minimization process in the term $\min_{\{\mathcal{B}_i\}} [S(\rho^{AB}|\{\mathcal{B}_i\})]$.
Fortunately, the quantum discord of Eq. (\ref{densitymatrix}) can be calculated
with the help of the results of \cite{Luo2008}
\begin{eqnarray}
\mathcal{D}(\rho^{AB})&=&\frac{1}{4}[(1-d_1-d_2-d_3)\log_2(1-d_1-d_2-d_3)\nonumber\\
&&+(1-d_1+d_2+d_3)\log_2(1-d_1+d_2+d_3)]\nonumber\\
&&+(1+d_1-d_2+d_3)\log_2(1+d_1-d_2+d_3)]\nonumber\\
&&+(1+d_1+d_2-d_3)\log_2(1+d_1+d_2-d_3)]\nonumber\\
&&-\frac{1-d}{2}\log_2{\frac{1-d}{2}}-\frac{1+d}{2}\log_2{\frac{1+d}{2}},\nonumber\\
d_1&=&d_2= p |f(t)|^2, d_3=-p, \nonumber\\
d&=&\max{\{|d_1|,|d_2|,|d_3|\}}.
\end{eqnarray}

In order to investigate the entanglement of two-qubit systems, we adopt the
entanglement measure concurrence introduced in \cite{Wootters1998}
\begin{equation}
C=\max {\{0,\chi _{1}-\chi _{2}-\chi _{3}-\chi _{4}\}},
\end{equation}
where $\chi _{i}$ ($i=1,2,3,4$) are the square roots of the
eigenvalues in decreasing order of the magnitude of the \textquotedblleft
spin-flipped" density matrix operator $R=\rho (\sigma _{y}\otimes \sigma
_{y})\rho ^{\ast }(\sigma _{y}\otimes \sigma _{y})$ and $\sigma _{y}$ is the
Pauli Y matrix, i.e., $\sigma _{y}=\left(
\begin{array}{cc}
0 & -i \\
i & 0%
\end{array}%
\right) $.
Concurrence of a quantum state ranges from 0, which corresponds
to an unentangled state, to 1, which corresponds to a maximally entangled state.
The concurrence of the above state is
\begin{eqnarray}
C(t)=\max{\{0,p|f(t)|^2-\frac{1-p}{2}\}}.
\end{eqnarray}
We will use this equation to calculate the
entanglement of the quantum system presented in this work.

\section{Results and discussions}

\subsection{Sudden death of quantum discord and entanglement}

\begin{figure}
\centering { \scalebox{1}[0.8]{\includegraphics{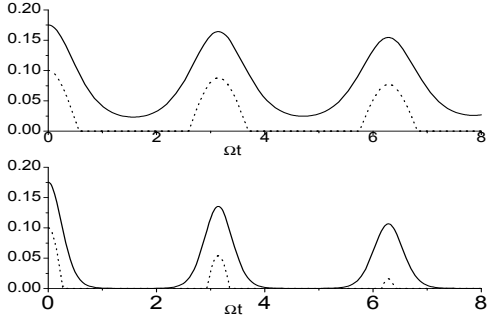}}}
\caption{Quantum discord (solid line) and entanglement (dotted line)
of two atoms are plotted as functions of the dimensionless scaled time
$\Omega t$ with $\gamma/\Omega=0.01$ and $p=0.4$.
Upper panel: $\alpha=0.5$. Lower panel: $\alpha=1$.
There is no DSD in the upper panel with $\alpha=0.5$. However, the
lower panel clearly shows the existence of DSD in the case of $\alpha=1$.}\label{fig1}
\end{figure}

We now want to investigate the dynamics of quantum discord and
entanglement. In Fig. \ref{fig1}, quantum discord and
entanglement are plotted as functions of the dimensionless scaled time
$\Omega t$ for $\alpha=0.5$ (upper panel) and $\alpha=1$ (lower panel).
Clearly, there is ESD in the present model as one can see from Fig. \ref{fig1}.
From the upper panel of Fig. \ref{fig1}, one may conclude that there is
no DSD, which is consistent with the results of \cite{Werlang2009,Wang2010}.
However, this is not always correct.  As one can easily observe from
the lower panel of Fig. \ref{fig1}, if we increase the intensity of
the coherent fields that is proportional to $|\alpha|^2$, there is DSD.
We note that in this figure, quantum discord is
larger than entanglement, which is coincidence with the observations
of \cite{Werlang2009,Wang2010}. However, this is not a general result
since quantum discord and entanglement are two different quantities and
there is no simple relative ordering between them.
For example, for the Werner states defined by
 $\rho=p|\psi_-\rangle\langle\psi_-|+\frac{1-p}{4}$,
where $|\psi_-\rangle=(|eg\rangle-|ge\rangle)/\sqrt{2}$, quantum discord
may be lager or smaller than entanglement \cite{Luo2008}.

We now want to explain why there is no DSD in the case of
dephasing channel in \cite{Werlang2009}. At first sight,
the elements of the density matrix of the dephasing case is
very similar to the matrix elements of Eq. (\ref{densitymatrix}).
Let us focus on the off-diagonal elements $\rho_{23}(t)$ of \cite{Werlang2009},
i.e., $\rho_{23}(t)=e^{-\Gamma t}\rho_{23}(0)$, where $\Gamma$ is the decay rate.
Obviously, the term $e^{-\Gamma t}$ becomes zero only in the asymptotic limit $t\rightarrow \infty$.
Quantum discord vanishes only in the asymptotic limit, which
behaves similarly to decoherence of each atom \cite{Werlang2009}.
In the present work, the off-diagonal element $\rho_{23}(t)$
is much more complicated than that of \cite{Werlang2009}
and it is possible for quantum discord to stay
zero for a finite time. Physically, the influence of
the interactions between atoms and cavities, and the mean photon number
and decay rate of dissipative cavities upon quantum discord has
not bee considered in \cite{Werlang2009}. Here, all the above influence upon the dynamics of
quantum discord and entanglement is taken into accounted.
In this sense, the results of \cite{Werlang2009} can only reveal
parts of the properties of quantum discord (no DSD), which is consistent with the upper panel of
Fig. \ref{fig1}. However, the lower panel of Fig. \ref{fig1}, which indicates the existence of
DSD, can not appear in \cite{Werlang2009}.

\subsection{Long-lived quantum discord and entanglement}
As we have pointed out previously, little attention has been
paid to long-lived quantum discord even though lots of work
has been made on long-lived entanglement \cite{Yu2006,Xu2005,Aty2006,Dajka2007,Aty2008}.
Here, we consider the question of whether there is long-lived quantum discord
and entanglement in the present model. The influence of the purity of the initial state
of atoms upon the long-time behavior of quantum discord and entanglement is also
discussed.

\begin{figure}
\centering { \scalebox{1}[0.8]{\includegraphics{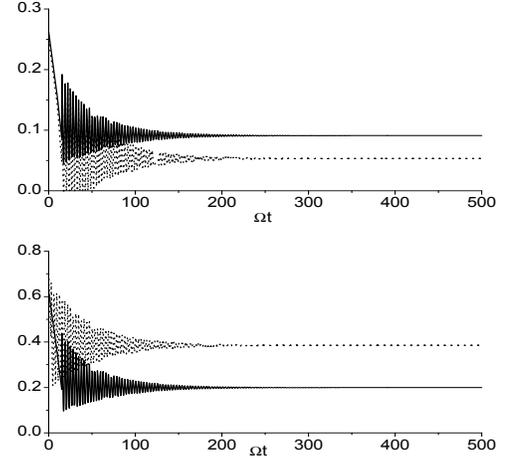}}}
\caption{Quantum discord (solid line) and entanglement (dotted line)
of two atoms are plotted as functions of the dimensionless scaled time
$\Omega t$ with $\gamma/\Omega=0.01$,
 and $\alpha=0.5$.
Upper panel: $p=0.5$. Lower panel: $p=0.8$.
Comparing the upper and lower panel, we see that long-lived quantum
discord could be larger (upper panel) or smaller (lower panel)
than long-lived entanglement. In other words, quantum entanglement may be
more robust than quantum discord in the presence of decoherence.
This result is different from the previous work\cite{Wang2010,Werlang2009}}\label{fig2}
\end{figure}

\begin{figure}
\centering { \scalebox{0.8}[0.8]{\includegraphics{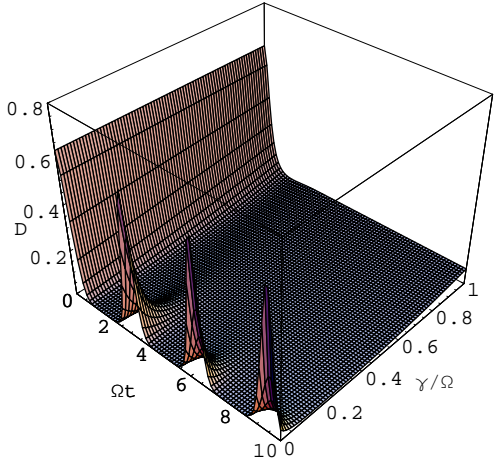}}}
\caption{Quantum discord of two atoms are plotted
as functions of the dimensionless scaled time $\Omega t$ and parameter $\gamma/\Omega$
with $\alpha=1$ and $p=0.8$.}\label{fig3}
\end{figure}

In Fig. \ref{fig2}, quantum discord and entanglement as functions of the
dimensionless scaled time $\Omega t$ are plotted for $p=0.5$ (upper panel)
and $p=0.8$ (lower panel). It is easy to observe that the long-time behavior
of quantum discord and entanglement depends on the purity of the initial state
of atoms. This figure is a direct evidence of the presence of long-lived quantum
discord and entanglement in the present system.
 Comparing the upper panel with
the lower one of Fig. \ref{fig2}, one can see that the amount of
long-lived quantum discord and entanglement will increase with the
increase of the parameter $p$. The relative ordering of quantum discord and
entanglement depends heavily on the purity $p$. For example,
in the case of $p=0.5$, long-lived quantum discord is larger than
long-lived entanglement. However, in the case of $p=0.8$, we
find that, in contrast to the results of \cite{Werlang2009,Wang2010},
the amount of long-lived quantum discord could be smaller
than that of long-lived entanglement. In other words,
quantum entanglement could be more robust than discord against decoherence.
Thus, it is difficult for us to make a general statement whether or not
quantum discord is more robust against decoherence than entanglement.
Intuitively, quantum discord is different from entanglement and it is
difficult to make a general conclusion about the relative ordering of
the amount of quantum discord and entanglement in the presence of decoherence.
In order to show the influence of decoherence of cavities upon the
quantum discord of two atoms, we plot the quantum discord of two atoms in Fig.3.
Note that it has been proved that white noise of
cavity fields can play a constructive role in the generation of
entanglement in cavity QED systems \cite{Plenio2002}. Here,
Fig. \ref{fig3} demonstrates that the dissipation
of cavity fields may also play a constructive role in the generation of
quantum discord. Note that in Figs. (\ref{fig1})-(\ref{fig3}), we have assumed
the atoms are initially prepared in $\rho_\phi$.
One can also consider the dynamics of quantum discord and entanglement if
the initial state of atoms is $\rho_\psi$ and the results are similar.

\section{Conclusions}
In the present work, we have studied the dynamics of quantum discord and entanglement
of two two-level atoms without direct interactions. Each atom is put into
a spatially separated and dissipative cavity in the dispersive limit.
We first investigated the short-time behavior of quantum discord and entanglement
in the presence of dissipation.
The amount of quantum discord and entanglement of two atoms decreases
with time if the interaction time is not very long.
Particularly, we have shown that both DSD and ESD could appear simultaneously in the present system.
This is different from the results of \cite{Werlang2009,Wang2010}.
Then, we discussed the long-time behavior of quantum discord and entanglement of
two atoms. We show there is long-lived quantum discord and entanglement
in the presence of the dissipation of cavities.
Our results show that quantum entanglement may
be more robust against decoherence than quantum discord.
Finally, we would like to point out that it could be possible to
study quantum discord of any dimensional bipartite states
from a geometrical point of view \cite{Dakic2010}.
It is also interesting to compare the results of any dimensional bipartite states
in the presence of decoherence with our work.

\acknowledgments
 This project is supported by the National¡¡
 Natural Science Foundation of China (Grant Nos 11047115 and 11065007),
 the Scientific Research Foundation of Jiangxi
Provincial Department of Education (Grant Nos GJJ10135 and GJJ09504), and the
Foundation of Talent of Jinggang of Jiangxi Province (Grant No
2008DQ00400).


\begin{thebibliography}{99}

\bibitem{Nielsen2000}
\Name{ Nielsen M. A. \and Chuang I. L.}
  \Book{Quantum
Computation and Quantum Information}
  \Publ{Cambridge University Press, Cambridge}
  \Year{2000}.

\bibitem{Raimond2001}
  \Name{Raimond J. M., Brune M. \and Haroche S.}
  \REVIEW{Rev. Mod.
Phys.}{73}{2001}{565}.

\bibitem{Chen2010}
  \Name{Chen L., Chitambar E., Duan R. Y., Ji Z. F. \and Winter A.}
  \REVIEW{Phys. Rev. Lett.}{105}{2010}{200501}.

\bibitem{Yu2004}
  \Name{Yu T. \and Eberly J. H.}
  \REVIEW{Phys. Rev. Lett.}{93}{2004}{140404}.

\bibitem{Yu2009}
  \Name{Yu T. \and Eberly J. H.}
  \REVIEW{Science (London)}{323}{2009}{598}.

\bibitem{Bellomo2008}
  \Name{Bellomo B., Lo Franco R. \and Compagno G.}
  \REVIEW{Phys. Rev.
A}{77}{2008}{032342}.

\bibitem{Rau2008}
  \Name{Rau A. R. P., Ali M. \and Alber G.}
  \REVIEW{Europhys. Lett.}{82}{2008}{4002}.

\bibitem{Zhang2009}
  \Name{Zhang J. S., Xu J.B. \and Lin Q.}
  \REVIEW{Eur. Phys. J. D}{51}{2009}{283}.

\bibitem{Jamroz2006}
  \Name{Jamr\'{o}z A.}
  \REVIEW{J. Phys. A: Math. Theor. }{39}{2006}{7727}.

\bibitem{Almeida2007}
  \Name{Almeida M. P., deMelo F., Hor-Meyll M., Salles A., Walborn S. P., Ribeiro P. H. S. \and Davidovich L.}
  \REVIEW{Science (London)}{316}{2007}{579}.

\bibitem{Salles2008}
  \Name{Salles A., deMelo F., Almeida M. P., Hor-Meyll M., Walborn S. P.  S.,  Ribeiro P. H. S. \and Davidovich L.}
  \REVIEW{Phys. Rev. A}{78}{2008}{022322}.

\bibitem{Laurat2007}
  \Name{ Laurat J., Choi K. S., Deng H., Chou C. W. \and Kimble H. J.}
  \REVIEW{Phys. Rev. Lett.}{99}{2007}{180504}.

\bibitem{Bennett1999}
  \Name{Bennett C. H., DiVincenzo D. P., Fuchs C. A., Mor T., Rains E., Shor P. W., Smolin J. A. \and Wootters W. K.}
  \REVIEW{Phys. Rev. A}{59}{1999}{1070}.

\bibitem{Horodecki2005}
  \Name{Horodecki  M., Horodecki P., Horodecki R., Oppenheim J., Sen A., Sen U. \and Synak-Radtke B.}
  \REVIEW{Phys. Rev. A}{71}{2005}{062307}.

\bibitem{Niset2006}
  \Name{Niset J. \and Cerf N. J.}
  \REVIEW{Phys. Rev. A}{74}{2006}{052103}.

\bibitem{Braunstein1999}
  \Name{Braunstein S. L., Caves C. M., Jozsa R., Linden N., Popescu S. \and Schack R.}
  \REVIEW{Phys. Rev. Lett.}{83}{1999}{1054}.

\bibitem{Meyer2000}
  \Name{Meyer D. A.}
  \REVIEW{Phys. Rev. Lett.}{85}{2000}{2014}.

\bibitem{Datta2005}
  \Name{Datta A., Flammia S. T. \and Caves C. M.}
  \REVIEW{Phys. Rev. A}{72}{2005}{042316}.

\bibitem{Datta2007}
  \Name{Datta A. \and Vidal G.}
  \REVIEW{Phys. Rev. A}{75}{2007}{042310}.

\bibitem{Datta2008}
  \Name{Datta A., Shaji A. \and Caves C. M.}
  \REVIEW{Phys. Rev. Lett.}{100}{2008}{050502}.

\bibitem{Dillenschneider2008}
  \Name{Dillenschneider R.}
  \REVIEW{Phys. Rev. B}{78}{2008}{224413}.

\bibitem{Sarandy2009}
  \Name{Sarandy M. S.}
  \REVIEW{Phys. Rev. A}{80}{2009}{022108}.

\bibitem{Cui2010}
  \Name{Cui J. \and Fan H.}
  \REVIEW{J. Phys. A: Math. Theor.}{43}{2010}{045305}.

\bibitem{Lanyon2008}
  \Name{Lanyon  B. P., Barbieri M., Almeida M. P.\and White A. G.}
  \REVIEW{Phys. Rev. Lett.}{101}{2008}{200501}.

\bibitem{Ollivier2001}
  \Name{Ollivier H. \and Zurek W. H.}
  \REVIEW{Phys. Rev. Lett.}{88}{2001}{017901}.

\bibitem{Henderson2001}
  \Name{Henderson L. \and  Vedral V.}
  \REVIEW{J. Phys. A: Math. Theor.}{34}{2001}{6899}.

\bibitem{Wang2010}
  \Name{Wang B., Xu Z. Y., Chen Z. Q.  \and Feng M.}
  \REVIEW{Phys. Rev. A}{81}{2010}{014101}.

\bibitem{Auyuanet2010}
  \Name{Auyuanet A. \and Davidovich L.}
  \REVIEW{Phys. Rev. A}{82}{2010}{032112}.

\bibitem{Sun2010}
  \Name{Sun Z. Y., Li L., Yao K. L., Du G. H., Liu J. W., Luo B., Li N. \and Li H. N.}
  \REVIEW{Phys. Rev. A}{82}{2010}{032310}.

\bibitem{Werlang2009}
  \Name{Werlang T., Souza S., Fanchini F. F.  \and Boas C. J. V.}
  \REVIEW{Phys. Rev. A}{80}{2009}{024103}.

\bibitem{Yu2006}
  \Name{Yu T. \and Eberly J. H.}
  \REVIEW{Opt. Commun.}{264}{2006}{393}.

\bibitem{Xu2005}
  \Name{Xu J. B.  \and Li S. B.}
  \REVIEW{New J. Phys.}{7}{2005}{72}.

\bibitem{Aty2006}
  \Name{Abdel-Aty M. \and Moya-Cessa H.}
  \REVIEW{Phys. Lett. A}{369}{2007}{372}.

\bibitem{Dajka2007}
  \Name{Dajka J., Mierzejewski M. \and Luczka J.}
  \REVIEW{J. Phys. A:
Math. Theor.}{40}{2007}{F879}.

\bibitem{Aty2008}
  \Name{Abdel-Aty M.}
  \REVIEW{ J. Phys. A: Math. Theor. }{41}{2008}{185304}.

\bibitem{Luo2008}
  \Name{Luo S.}
  \REVIEW{Phys. Rev. A}{77}{2008}{042303}.

\bibitem{Scully1997}
\Name{Scully M. \and Zubairy M. S.}
  \Book{Quantum Optics}
  \Publ{Cambridge University Press, Cambridge}
  \Year{1997}.


\bibitem{Meystre1992}
  \Name{Meystre P.}
  \REVIEW{Phys. Rep.}{219}{1992}{243}.

\bibitem{Zhang2010}
  \Name{Zhang J. S., Chen A. X. \and Abdel-Aty M.}
  \REVIEW{J. Phys. B: At. Mol. Opt.
Phys.}{43}{2010}{ 025501}.

\bibitem{Zhang20091}
  \Name{Zhang J. S.\and Xu J. B.}
  \REVIEW{Opt. Commun.}{282}{2009}{2543}.

\bibitem{Zhang20092}
  \Name{Zhang J. S.\and Xu J. B.}
  \REVIEW{Can. J. Phys.}{87}{2009}{1031}.

\bibitem{Wootters1998}
  \Name{Wootters W. K.}
  \REVIEW{Phys. Rev. Lett.}{80}{1998}{2245}.

\bibitem{Plenio2002}
  \Name{Plenio M. B. \and Huelga S. F.}
  \REVIEW{Phys. Rev. Lett.}{88}{2002}{197901}.

\bibitem{Dakic2010}
  \Name{Dakic B., Vedral V. \and Brukner C.}
  \REVIEW{ Phys. Rev. Lett.}{105}{2010}{190502}.
\end{thebibliography}
\end{document}